\documentclass{article}

\usepackage{graphicx}
\usepackage{amsmath}
\usepackage{amssymb}

\begin{document}

\title{Anisotropic elastic theory\\ of preloaded granular media}
\author{Cyprien Gay$^{1}$ and Rava A.~da~Silveira$^{2}$ \\~\\
$^{1}$Centre de recherche Paul-Pascal--CNRS,\\ 
Av. Schweitzer, 33600 Pessac, France,\\ 
$^{2}$ Lyman Laboratory of Physics, Harvard University,\\
Cambridge, Massachusetts 02138, U.S.A.\\
}
\maketitle

{\em45.70.Cc} {Static sandpiles; granular compaction}\newline
{\em 46.25.-y} {Static elasticity}\newline
{\em 62.20.Dc} {Elasticity, elastic constants}

\begin{abstract}
A macroscopic elastic description of stresses in static, preloaded granular
media is derived systematically from the microscopic elasticity of
individual inter-grain contacts. The assumed preloaded state and friction at
contacts ensure that the network of inter-grain contacts is not altered by
small perturbations. The texture of this network, set by the preparation of
the system, is encoded in second and fourth order fabric tensors. A small
perturbation generates both normal and tangential inter-grain forces, the
latter causing grains to reorient. This reorientation response and the
incremental stress are expressed in terms of the macroscopic strain.
\end{abstract}

The transmission of stress in granular media has a rich phenomenology \cite%
{reviews,goddard,pgg}, as illustrated by the emblematic sand pile problem.
In a conical pile obtained by pouring grains from a point source (hopper
outlet), the pressure profile at the base of the pile is not proportional to
the height of the pile, as it would if the weight of each grain were
transmitted strictly vertically; nor does the pressure vary monotonically
from edge to center, as predicted by traditional isotropic elastic or
elasto-plastic approaches. Rather, the pressure profile develops a local
minimum (often termed `pressure dip') at the center of the pile base, below
the apex of the pile \cite{liu,brockbank,vanel,geng2001}. By contrast, if a
pile of the same shape is prepared layer by layer, \textit{e.g.}, by
sprinkling from an extended source, the pressure profile \textit{does}
acquire a maximum at the center of the base (`pressure hump') \cite%
{liu,brockbank,vanel,geng2001}. These and similar experiments indicate that
the local structure of the pile (often called `texture'), which governs
stress transmission, depends sensitively on the preparation of the system.
In two dimensional packings of monodisperse disks~ \cite{geng2001}, for
example, the observed distribution of inter-grain contact orientations
reflects the local grain ordering, with a clear six-fold modulation.
However, as expected from simple symmetry arguments (see Fig.~\ref%
{preparation} for an illustration), the distribution displays a stronger
degree of anisotropy (including two- and four-fold modulations) in a usual
pile, built with a point source, than in a pile grown by sprinkling grains.
Further experiments~\cite{geng2001,geng}, as well as simulations~\cite%
{goldenberg}, confirm that the anisotropy of the \textit{microscopic}
contact distribution is correlated with the presence of a \textit{macroscopic%
} pressure dip.

As it proved difficult to fit the wealth of observed stress profiles~\cite%
{liu,brockbank,vanel,geng2001,clement,reydellet,geng} with usual elastic
theories, in the past decade physicists introduced highly idealized discrete
models of static granular media~\cite%
{liu,coppersmith,bouchaud1995,claudin1998} based on probabilistic rules of
force transmission between neighbouring grains. Although these models
reproduce experimental features, they leave stresses underspecified (and
consequently rely on a somewhat \textit{ad hoc} constraint among stress
components), as a result of an incomplete treatment of inter-grain forces.
Furthermore, careful investigations of force transmission models \cite%
{bouchaud2001,socolar2002,socolar2003} seem to indicate that disorder in the
granular packing might cause it to behave again as an elastic medium at
large scales. In this spirit, Ref.~\cite{otto} reviews the various possible
outcomes of an (orthotropic) \cite{green-taylor} anisotropic elastic theory 
\cite{green-zerna}, without however addressing the origin of the anisotropy
in the context of granular media.

In a continuum, any volume element is compactly embedded in the surrounding
medium. By contrast, in more fragile granular media, stress is transmitted
through isolated inter-grain contacts, and under a perturbation grains
slightly reorient with respect to their surroundings. This occurs even in
the case of a simple compressive strain: consider for example a compression
along the $x$-axis and focus on the contact of a given grain with a
neighbouring grain. The torque that this contact imparts to the grain
vanishes only if the contact is parallel or perpendicular to the $x$-axis.
Thus, generically, a non-vanishing total torque is applied on a grain by
contacting grains, and consequently it reorients so as to restore the torque
to zero in equilibrium. A continuum theory derived from inter-grain forces
must include effects of grain reorientation. In this vein, a generalization
of usual elasticity that accommodates the `micro-rotation'\ of points in
addition to the `macro-rotation'\ of the medium was formulated by the
Cosserat brothers~\cite{cosserat}, and subsequently generalized and applied
to `complex' continuum media \cite{lakes,chang95}.

In the present work, our starting point is the well-established elastic
theory of inter-grain contacts~\cite{hertz,johnson}; on this basis, we
construct a specific macroscopic elastic theory \cite{chang-note} that
encodes the texture of the granular network. We consider a preloaded
granular medium~\cite{preloaded} and small applied incremental stresses, so
that the corresponding elastic response be linear. While the unperturbed
inter-grain forces reflect the rolling and sliding~\cite{radjai} that occur
during the preparation of the system, as well as the external preloading, if
any, we expect the response of the medium to small incremental stresses to
be sensitive only to properties of the (unperturbed) contact network. Below,
we derive general expressions for the (anisotropic) elastic linear response
and the associated grain reorientation field.

\begin{figure}[tbp]
\center
\includegraphics[width=0.5\linewidth]{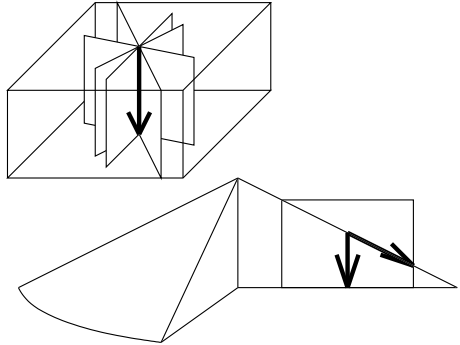}
\caption{In a pile constructed layer by layer (top), the local texture of
the packing has only one prefered (vertical) direction and is therefore
statistically symmetric about any vertical plane (and hence invariant under
rotation about a vertical axis). By contrast, in a usual pile grown with a
point source (bottom), the presence of another special direction (the
downhill direction) lowers the symmetry, and the local texture is
statistically symmetric with respect to a single vertical plane, that which
contains the downhill direction.}
\label{preparation}    
\end{figure}

\begin{figure}[tbp]
\center
\includegraphics[width=0.5\linewidth]{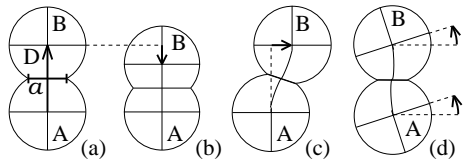}
\label{hertz-thick}
\caption{Schematic illustration of the preloading and deformation of two
contacting grains. (a) The diameter $a$ of the contact region between two
preloaded grains is non-vanishing (over-emphasized on the figure). A
restoring force about this compressed state is generated by (b) further
compression (or traction), (c) shear, or (d) reorientation. If $a\ll D$, (c)
and (d) are equivalent for small perturbations. Correspondingly, tensors $%
\protect\varepsilon$ and $\Omega$ enter Eq.~({\protect\ref{hooke}}) on the
same footing.}
\end{figure}

The contacting region between two preloaded spherical grains A and B~(Fig.~%
\ref{hertz-thick}(a)) is a disk with diameter $a\approx \lbrack 3FD(1-\nu _{%
\text{g}}^{2})/8E]^{1/3}$, where $F$ is the inter-grain compressive force, $%
D $ the grain diameter (or, equivalently, the centre-to-centre distance), $E$
the Young modulus and $\nu _{\text{g}}$ the Poisson ratio of the grain
constitutive material \cite{hertz}. Because of the high friction established
by the preloading, essentially no slip occurs at the contact upon a
perturbation of the grains \cite{friction}. Consequently, the linear elastic
response~of such a contact is characterized by stiffness constants $%
k_{\parallel }$ for compression or traction along the centre-to-centre
vector $\vec{D}$ (henceforth called \textit{contact vector}) (Fig.~\ref%
{hertz-thick}(b)) and $k_{\perp }$ for transverse forces due to shearing
(Fig.~\ref{hertz-thick}(c)) \cite{johnson}. Transverse forces are also
generated by the reorientation of two contacting grains (Fig.~\ref%
{hertz-thick}(d)). Typically, $a\ll D$, implying $k_{\parallel },k_{\perp
}\sim Ea$, and the ratio $k_{\perp }/k_{\parallel }$ is fixed by $\nu _{%
\text{g}}$; for example, for a purely compressively preloaded contact $%
k_{\bot }/k_{\Vert }=(2-2\nu _{\text{g}})/(2-\nu _{\text{g}})$~\cite{johnson}%
. In addition to compressive and shearing forces, torques may be transmitted
through twisted inter-grain contacts, with a corresponding stiffness $%
k_{t}\sim Ea^{3}$ relating the twist angle to the torque. As long as $a\ll D$%
, such twisting torques contribute negligibly to the behaviour of the medium
and we neglect them in the present work. (In the case of non-spherical (
\textit{e.g.}, faceted) grains, in which twisting modes may be more
relevant, one can keep track of them by defining a torque flux tensor, often
termed `couple stress' in Cosserat theories \cite{cosserat,lakes}, and
proceed as with the stress tensor defined below.) Combining these various
elements, we write the force exerted by grain B on grain A as 
\begin{equation}
f_{i}^{\text{AB}}=k_{ij}^{\text{AB}}\left( u_{j}^{\text{B}}-u_{j}^{\text{A}%
}\right) -k_{ij}^{\text{AB}}\frac{\Omega _{jk}^{\text{A}}+\Omega _{jk}^{%
\text{B}}}{2}D_{k}^{\text{AB}},  \label{force}
\end{equation}%
where%
\begin{equation}
k_{ij}^{\text{AB}}=k_{\parallel }\,\frac{D_{i}^{\text{AB}}D_{j}^{\text{AB}}}{%
D^{2}}+k_{\perp }\left( \delta _{ij}-\frac{D_{i}^{\text{AB}}D_{j}^{\text{AB}}%
}{D^{2}}\right)
\end{equation}%
and the sum over repeated indices is understood. The tensor $\Omega ^{\text{G%
}}$ is antisymmetric and is defined such that the rotation matrix $\left[
\exp \left( \Omega ^{\text{G}}\right) \right] _{ij}\approx \delta
_{ij}+\Omega _{ij}^{\text{G}}$ describes the \textit{reorientation} of grain
G. We note that the mean reorientation, $(\Omega ^{\text{A}}+\Omega ^{\text{B%
}})/2$, results in a sheared contact (Fig.~\ref{hertz-thick}(d)) and thus
contributes to the force (Eq. (\ref{force})), while the difference $\Omega ^{%
\text{A}}-\Omega ^{\text{B}}$ corresponds to grain rolling, which does not
affect the force between spherical grains \cite{loops}.

In order to build a continuum description of a preloaded granular medium, we
identify the displacements $\vec{u}^{\text{G}}$ of the centers of the grains
(labelled by G) with a smoothly varying, macroscopic displacement field $%
\vec{u}(\vec{r})$. Similarly, we identify the grain reorientation $\Omega ^{%
\text{G}}$ with a smooth reorientation field $\Omega (\vec{r})$, to be
determined. In doing so, we neglect the non-affine component of the grain
displacements. This approximation is exact in the limit of a simple
crystalline granular network, in which all grains are equivalent, but we
expect it to be good even away from this limit. If the packing disorder
becomes too important, the medium can still be described by an elastic
theory, although its effective stiffness constants may depart significantly
from those obtained below~\cite{tanguy}. Grain displacements can now be
expressed in terms of the strain tensor, as $u_{i}^{\text{B}}-u_{i}^{\text{A}%
}=\varepsilon _{ij}D_{j}^{\text{AB}}$, where $\varepsilon _{jk}=\left(
\partial _{j}u_{k}+\partial _{k}u_{j}\right) /2$ as usual, and the force
transmitted through a contact whose contact vector lies along the direction $%
\alpha $ reads%
\begin{equation}
f_{i}(\alpha )=k_{ij}(\alpha )\left( \varepsilon _{jk}-\Omega _{jk}\right)
D_{k}(\alpha ),  \label{hooke}
\end{equation}%
where $k_{ij}(\alpha )$ and $D_{k}(\alpha )$ refer to a contact normal to
the direction $\alpha $. We emphasize that, as it appears explicitly in Eq.~(%
\ref{hooke}), the grain reorientation $\Omega $ affects inter-grain forces;
it does \textit{not} represent the antisymmetric part of the displacement
gradient (which corresponds to solid body rotation and does not generate
stress).

For a description in terms of stresses, it is convenient to define the
average number $\mu (\alpha )\;\mathrm{d}\alpha $ of contact vectors that
lie within a solid angle $\mathrm{d}\alpha $ about direction $\alpha $, per
unit volume, and to introduce local \textit{fabric tensors}~\cite{goddard} $%
Q $ and $P$ whose slow spatial variations reflect those of $D$, $%
k_{\parallel } $, $k_{\perp }$, and $\mu $, as%
\begin{equation}
Q_{kj}=\int k_{\bot }D_{k}D_{j}\;\mu \;d\alpha  \label{fabric-tensor-Q}
\end{equation}%
and 
\begin{equation}
P_{ijkl}=\int (k_{\Vert }-k_{\bot })\frac{D_{i}D_{j}D_{k}D_{l}}{D^{2}}\;\mu
\;d\alpha .  \label{fabric-tensor-P}
\end{equation}%
In equilibrium, the total torque $\sum_{\text{G}^{\prime }}\vec{f}^{\text{GG}%
^{\prime }}\times \vec{D}^{\text{GG}^{\prime }}$ imparted to grain G by its
neighbours (labelled by G$^{\prime }$) vanishes. Locally averaged in space,
this condition becomes 
\begin{equation}
\int \vec{f}(\alpha )\times \vec{D}(\alpha )\mu (\alpha )\mathrm{d}\alpha =0
\end{equation}%
or, equivalently,%
\begin{equation}
\Omega Q+Q\Omega =\varepsilon Q-Q\varepsilon .  \label{commutator}
\end{equation}%
This identity will be useful in the derivation of the stress in terms of the
strain only, and confirms that grains reorient unless their contacts lie
along a principal axis of the strain (on average), \textit{i.e.}, unless $%
\varepsilon $ and $Q$ commute. In order to compute the stress, we note that $%
D_{j}(\alpha )\mu (\alpha )\mathrm{d}\alpha $ is the number of contact
vectors per unit surface area, oriented within $\mathrm{d}\alpha $ about
direction $\alpha $, that intersect a surface normal to direction $j$. Since 
$f_{i}(\alpha )$ is the $i$-th component of the force transmitted along such
contact vectors, the stress tensor can be written as%
\begin{equation}
\sigma _{ij}=\int \,f_{i}(\alpha )\,D_{j}(\alpha )\mu (\alpha )\mathrm{d}%
\alpha .  \label{stress}
\end{equation}%
While this expression is not symmetric in general, it is symmetrized as
expected by the above vanishing-torque condition (Eq. (\ref{commutator})).
Combining Eqs. (\ref{hooke}--\ref{stress}), we obtain%
\begin{equation}
\Omega =\int_{0}^{\infty }e^{-sQ}(\varepsilon Q-Q\varepsilon )e^{-sQ}\mathrm{%
d}s\text{ \ \ \ \ \ \ or \ \ \ \ \ \ }\Omega _{ij}=\frac{q_{j}-q_{i}}{%
q_{j}+q_{i}}\,\varepsilon _{ij},  \label{omega}
\end{equation}%
and%
\begin{equation}
\sigma =\int_{0}^{\infty }e^{-sQ}\;2Q\varepsilon Q\;e^{-sQ}\mathrm{d}%
s+P:\varepsilon \text{ \ \ \ \ \ \ or \ \ \ \ \ \ }\sigma _{ij}=2\frac{%
q_{i}q_{j}}{q_{i}+q_{j}}\,\varepsilon _{ij}+P_{ijkl}\varepsilon _{kl}.
\label{sigma}
\end{equation}%
The right-hand expressions refer to a basis in which $Q$ is diagonal, and $%
q_{i}$ is the $i$-th eigenvalue of $Q$; indices $i$ and $j$ are \textit{not}
summed over. As expected, the reorientation, $\Omega $, depends on $k_{\perp
}$ (through $Q$) but not on $k_{\parallel }$. Equation~(\ref{omega}) is
qualitatively corroborated by the experimental observation~\cite{calvetti}
that $\Omega $ is largest in regions of large deformation.

Equations~(\ref{omega},\ref{sigma}) constitute our central result and
describe the response of a preloaded granular medium to an incremental
strain. If the medium is statistically isotropic ($\mu =\rho /4\pi $
constant), we recover standard isotropic elasticity, with Lam\'{e}
coefficients $\lambda _{\text{L}}=\rho D^{2}(k_{\parallel }-k_{\perp })/15$
and $\mu _{\text{L}}=\rho D^{2}(k_{\parallel }/15+k_{\perp }/10)$, in
agreement with the results of Chang and Gao~\cite{chang95,chang-note}. In
particular, if the Poisson ratio $\nu _{\text{g}}$ of the grain constitutive
material is positive, that of the preloaded granular medium, $\nu =\nu _{%
\text{g}}/(10-6\nu _{\text{g}})$, lies between $0$ and $1/14$. (We recall,
however, that in the isotropic limit our coarse-grained theory is only
approximate). In an anisotropic medium, the fabric tensors $Q$ and $P$
encode the texture of the medium through the dependence of $D$, $%
k_{\parallel }$, $k_{\perp }$, and $\mu $ on $\alpha $. Indeed, if $\mu $
varies with $\alpha $ or if the preloading stress is anisotropic, stiffness
constants may also vary with $\alpha $ as contacts are then more or less
compressed depending on their orientations and may be sheared. The elastic
response described by Eq.~(\ref{sigma}) involves only the first few
multipolar components of the quantities $D^{2}k_{\perp }\mu $ (through $Q$)
and $D^{2}(k_{\parallel }-k_{\perp })\mu $ (through $P$). Monopoles
contribute to the isotropic part of the response. Dipoles are absent due to
the fore-aft symmetry of the contact distribution. Quadrupoles in $Q$ and $P$
and octupoles in $P$ contribute to the anisotropic elastic response. Higher
order multipoles play no role in the present linear theory. As to
reorientation, given by Eq.~(\ref{omega}), the response is affected solely
by the quadrupolar component of $D^{2}k_{\perp }\mu $.

Before concluding, we mention that it is possible to construct an iterative
scheme for calculating stresses in a sand pile. 
It relies on the assumption
that the mechanical noise due to avalanches of grains at the free surface,
and other factors such as temperature fluctuations, do not cause significant
rearrangements of the contact network~\cite{pgg}. A rigorous implementation
of an iterative scheme still requires knowledge of the complicated
mechanisms involved in the growth of the pile, which include both grain
avalanching~\cite{rajchenbach} and the non-trivial frictional mechanics
putting grains to rest~\cite{radjai,halsey}. Indeed, these determine both
the angular contact distribution and the tangential compressive stresses 
at the free surface.

As a first attempt at a quantitative understanding, one may focus on the
simpler problem of calculating the macroscopic response to an applied point
force~\cite{green-taylor,green-zerna,claudin1998,goldenberg,silveira,otto} 
for different possible textures of the granular medium. 
Using Eqs. (\ref{omega}, \ref{sigma}), 
we find that stress transmission is very sensitive to the texture
as summarized by tensors $Q$ and $P$, and comes in qualitatively diverse
forms. Specifically, depending on the degree of anisotropy, the incremental
stress profile may be single- or double-peaked~\cite{silveira}, in agreement
with Refs.~\cite{goldenberg,otto}; furthermore, the width of the peaks
depends also on the degree of anisotropy and, in the single-peak case, may
depart from its isotropic counterpart~\cite{silveira}.

In sum, we have derived a macroscopic formulation of stresses in granular
media. This formulation differs from earlier ones in that it incorporates
the possibly anisotropic texture of the granular network as well as the
reorientation of grains induced by macroscopic deformations. Moreover, the
central objects in our formulation, the fabric tensors, are defined in terms
of the microscopic parameters that characterize inter-grain contacts.

One can wonder whether this description might bear some signature of a
salient experimental fact: the existence of highly stressed regions known as
`force chains'~\cite{liu,clement,geng}, arranged in a percolating,
filamentous network that appears to convey a large fraction of the stress
through the medium. Analogously~\cite{witten,didonna} to force chains,
stresses in a crumpled elastic sheet are confined mostly within narrow
regions, the folds, if the system is allowed to bend in the third dimension
(of the embedding space)~\cite{witten,didonna}. In our case, grain
reorientation may be viewed as an additional degree of freedom which the
system uses to relieve stress. It would be interesting to investigate
whether this reorientational freedom of grains may favor, in a heterogeneous
system, stress condensation reminiscent of force chains.

\section*{Acknowledgments}

It is a pleasure to thank J.-P.~Bouchaud, E.~Cl\'{e}ment, D.S.~Fisher,
P.-G.~de Gennes, B.I.~Halperin, J.~Rajchenbach, J.R.~Rice, and J.-N.~Roux
for uselful discussions, and P.G.~de Gennes and T.A.~Witten for inspiring
lectures. We thank the referees for valuable comments. R.A.S. acknowledges
the support of the Harvard Society of Fellows, the Milton Fund, and the 
\textit{Fonds national suisse} through a Young Researcher Fellowship.


\begin{thebibliography}{99}
\bibitem{friction} A slight shearing of a preloaded contact causes slip only
at its periphery~\cite{johnson}. Since the contact area is barely affected,
the stiffness constants $k_{\parallel }$ and $k_{\perp }$ do not vary to
first order.
\end{thebibliography}
\end{document}